\documentclass[reprint,amsmath,amssymb,aps,pre,]{revtex4-2}
\usepackage[caption=false]{subfig}
\usepackage{listings}
\usepackage{xcolor}
\usepackage{multirow}
\usepackage{graphicx}
\usepackage{dcolumn}
\usepackage{bm}
\usepackage{braket}
\usepackage{hyperref}

\makeatletter
\newcommand{\vast}{\bBigg@{3}}
\newcommand{\Vast}{\bBigg@{4}}
\begin{document}
\title{Unusual crosstalk in coincidence measurement searches for quantum degeneracy}
\author{Arjun Krishnan U M}
\author{Raul Puente}
\author{M.A.H.B. Md Yusoff}
\author{Herman Batelaan}
 \email{hbatelaan@unl.edu}
\affiliation{Department of Physics and Astronomy, University of Nebraska-Lincoln, Jorgensen Hall, Theodore, 855 N 16th St 208, Lincoln, NE 68588, USA}
\begin{abstract}
A dip in coincidence peaks for an electron beam is an experimental signature to detect Coulomb repulsion and Pauli pressure. This paper discusses another effect that can produce a similar signature but that does not originate from the properties of the physical system under scrutiny. Instead, the detectors and electronics used to measure those coincidences suffer significantly even from weak crosstalk. A simple model that explains our experimental observations is given. Furthermore we provide an experimental approach to correct for this type of crosstalk.
\end{abstract}

\maketitle
\section{Introduction}
Consider two electrons emitted from a source that travel through free space to a detector that measures their arrival time. The electrons do not like to arrive together in time. This can be due to repulsive Coulomb forces between the electrons \cite{keramati2021,ropers2023,hommelhoff2023,luiten2024} or Pauli blockade. The latter is the Hanbury Brown and Twiss (HBT) antibunching in time \cite{hbt1954}, and its observation has been claimed twice for free electrons by Hasselbach in 2002 \cite{hasselbach2002} and Kuwahara in 2021 \cite{kuwahara2021,batelaan2021} amidst other searches \cite{kodama2011}. The experiments are done and report quantum degeneracies of ~$10^{-4}$. The reason for the low value is that the experiments were done using emission sources that either ran continuously from a nanotip or are generated in picosecond bunches from a micro-sized surface, and only rarely are two electrons emitted nearly simultaneously and close to each other in phase space. Given the small size of the effect, it is important to consider instrumental effects. It is well known that crosstalk can give rise to false coincidence signals. Hasselbach captures this succinctly: ``Capacitive crosstalk between the collectors was well below 1$\%$ and did not cause spurious coincidences.'' However, there is a second crosstalk mechanism that even in the absence of spurious coincidences still can lead to a false antibunching signal. 

In this paper we measure and model this signal and provide an experimental method to correct for this false dip in coincidence rate. The idea is that electromagnetic coupling between start and stop signal wires (or other hardware parts) causes an inverted crosstalk signal that diminishes the real signal. This diminishes the coincidence rate only when a start and stop signal are simultaneously present, thus forming a false antibunching dip. Surprisingly, we observe that $1\%$ crosstalk can lead to an 8$\%$ dip, which is significant in view of the sub 0.1$\%$ quantum degeneracy signals reported. 

\section{Experiment}
\begin{figure*}
     \includegraphics[width=15cm]{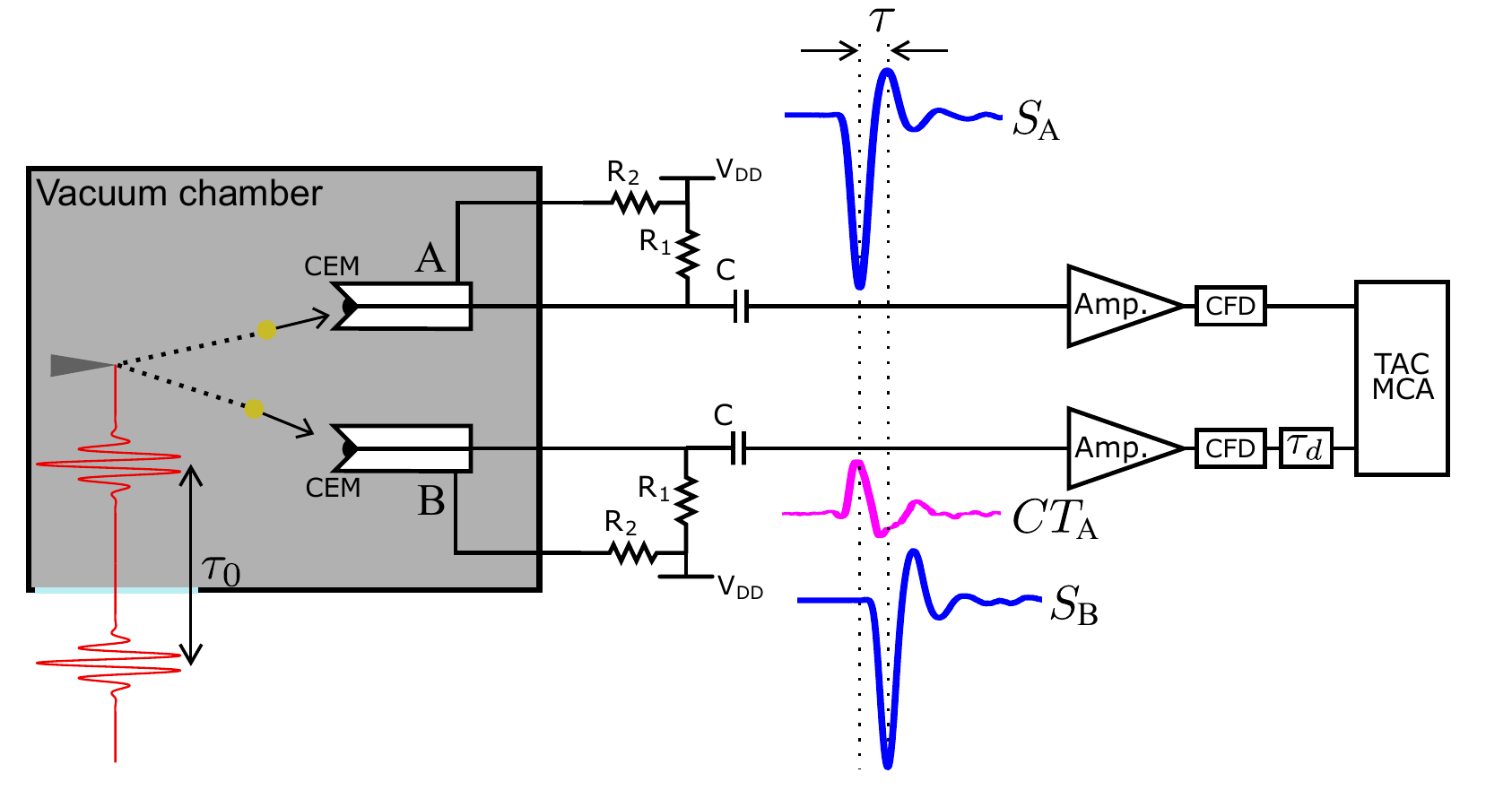}
     \caption{Schematic set-up. Laser pulses (red) generate electron pulses from a nanotip, which are detected in coincidence by channel electron multipliers (CEM). Recorded pulse traces, $S_{A/B}$, and the crosstalk of $S_{A}$, $CT_{A}$ are shown. Amplifiers (Amp.) and constant fraction discriminators (CFD) are connected to the start and stop of a time-to-amplitude-converter (TAC) and multichannel analyzer (MCA). A fixed delay $\tau_{d}$ controls the antibunching dip location.}
     \label{fig:schematic}
\end{figure*}
\begin{figure}
     \includegraphics[width=8.5cm]{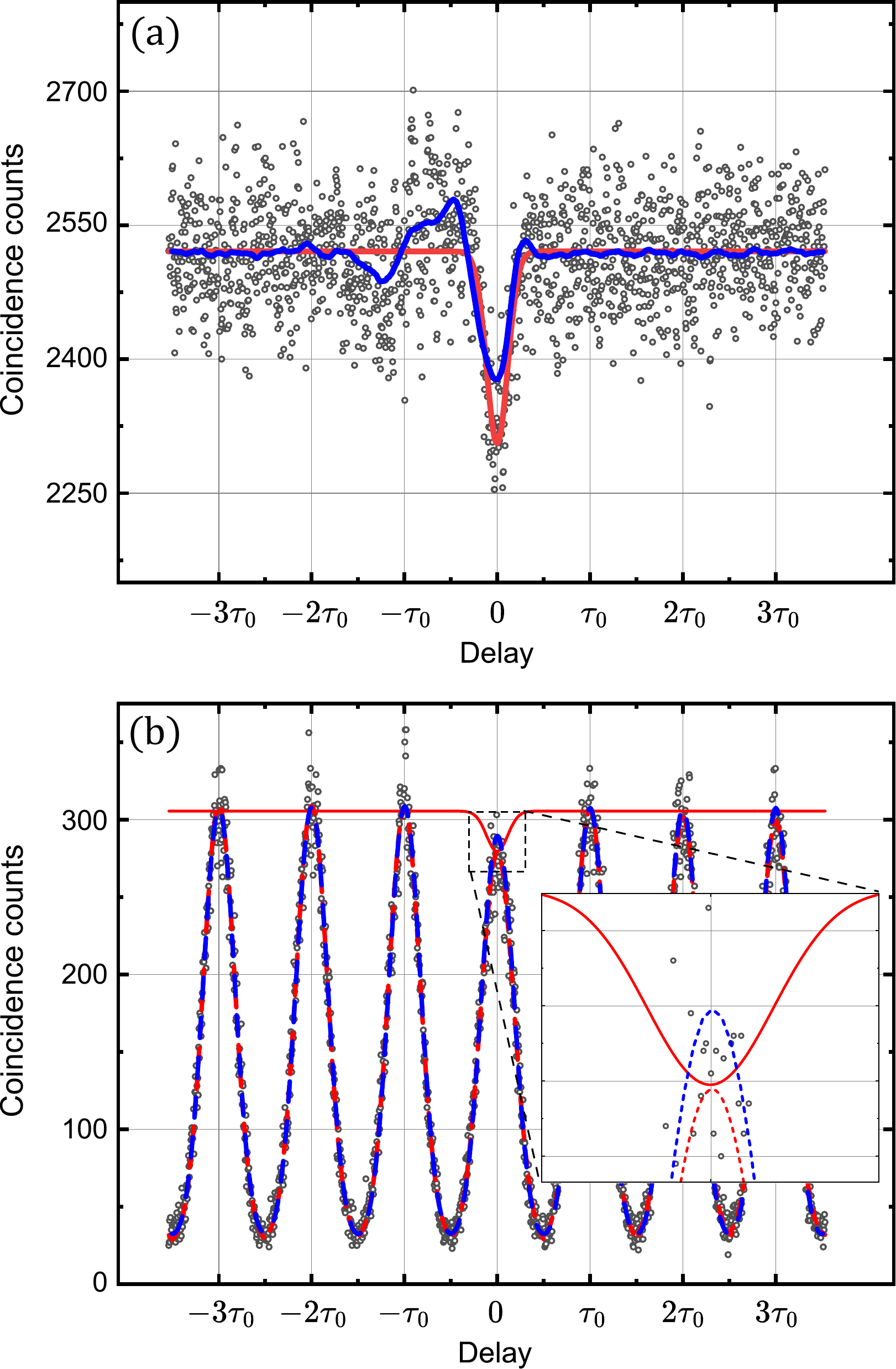}
     \caption{\textbf{(a)} The coincidence spectrum of electrons from an electrically heated tungsten wire source (black circles). The dip in coincidence counts at zero time delay due to crosstalk has been fitted as a Gaussian dip against a flat background (red line). The fitting indicates a $8.5 \%$ dip in coincidence counts at zero delay compared to the background. The blue line indicates the simulated coincidence spectrum for a continuous thermal source with a $6.1\%$ dip at zero time delay. \textbf{(b)} The coincidence spectrum of electrons from a tungsten nanotip illuminated by $800$ nm laser pulses with a pulse period $\tau_0 $ (black circles). The dashed red line represents the Gaussian fit of the peaks. The red line from the top panel is scaled so that the flat background matches the average maximum height of the non-central peaks. The dip in the red line in the bottom panel at $0\tau_0$ matches the maximum in the fit of the central peak indicating this effect is due to crosstalk. The dashed blue line is the simulated spectrum.}
     \label{fig:ZeroBounces}
\end{figure}

Fig.\ref{fig:schematic} shows the schematic of an experiment to measure the antibunching of electrons produced by a femtosecond pulsed laser. The laser pulse photoemits electrons from a tungsten nanotip source and an accelerating DC-voltage on the nanotip pushes them toward two channel electron multiplier (CEM) detectors A and B, referred to as start and stop detectors, respectively. The signal pulses from the detectors pass through a preamplifier which amplifies them from $\sim 10$ mV to $\sim 1$ V. The constant fraction discriminator (CFD) converts this into logical pulses if the pulse height is above a certain threshold value $V_{th}$. The time-to-amplitude converter (TAC) produces a pulse with a height proportional to the time delay between the arrival of two logical pulses between A and B. A multichannel analyzer (MCA) records a histogram of the number of pulses with these heights, producing the coincidence spectrum shown in Fig.\ref{fig:ZeroBounces}.

The coincidence spectrum consists of a series of peaks centered around integer multiples of the inverse repetition rate of the laser $\tau_0 = 11.8$ ns. Since the photoemission of electrons is an inherently random process, the number of electrons emitted by a single pulse, $m$, follows a Poisson distribution;
\begin{equation}
    P_m(\lambda)=\frac{\lambda^me^{-\lambda}}{m!}.
\end{equation}
Here, $\lambda$ is the average number of electrons emitted by a laser pulse. The coincidence counts at the peak centered around delay $n \tau_0$  ($n \neq 0$) is given by \cite{keramati2021},
\begin{equation}
    N(n\tau_0)=\epsilon_A\epsilon_B\left(1-e^{-\lambda}\right)^2e^{-(n-1)\epsilon_B\lambda} (N_p-n),
\end{equation}
where $\epsilon_A$ and $\epsilon_B$ denote the detector efficiencies. In this case, that includes the probability of the emitted electron to travel from the source to the detector. $N_p$ is the total number of laser pulses incident on the source during the measurement period. For zero time delay ($n=0$) the coincidence counts are produced by multiple electrons emitted by a single laser pulse and is given by,
\begin{equation}
 N(0\tau_0)=\epsilon_A\epsilon_B\lambda^2N_p.
\end{equation}

For small $\lambda$ and large $N_p$, the ratio of $N(0\tau_0)$ and $N(n\tau_0)$ reaches $1$ for relatively small $n$ values. In other words, we expect the peaks to have the same number of counts. A reduction in the central peak, $N(0\tau_0)$, is the sought-after signature indicative of interaction between the electrons – either due to Pauli repulsion or Coulomb repulsion. Note that for a continuous source this leads to a dip in a continuous background. In contrast, the dip in Fig.\ref{fig:ZeroBounces}a and the reduction of the central peak in Fig.\ref{fig:ZeroBounces}b are due to crosstalk, that we intentionally introduce and control.

To explain this false signal, consider the front end of the detection. CEM detectors work based on an avalanche process where an incident electron triggers a chain reaction of producing secondary electrons that are accelerated by fields to produce an electric pulse in the mV range. The rise time of the pulses is in the order of a few nanoseconds and can produce capacitive and/or inductive crosstalk in circuits when parts of the two arms of the detector are in close proximity to each other. The signal from CEM A is measured with an oscilloscope after capacitive decoupling from the high voltage, $V_{DD}$. The blue trace $S_{A}$ (Fig.\ref{fig:schematic}) is a 500X sample-averaged oscilloscope trace indicating the detection of a single electron at CEM A. The magenta trace $CT_{A}$ is the crosstalk signal (in the absence of an electron at CEM B) detected the same way and has about $\sim 1\%$ amplitude of $S_{A}$. When an electron is detected both at CEM A and B, the ``real'' signals and the crosstalk signal are overlapping for zero delay coincidences. Note that the signal has an inverted shape. This leads to a reduction in pulse height and a lower count rate after the discriminator.

\begin{figure*}
     \centering
     \includegraphics[width=\textwidth]{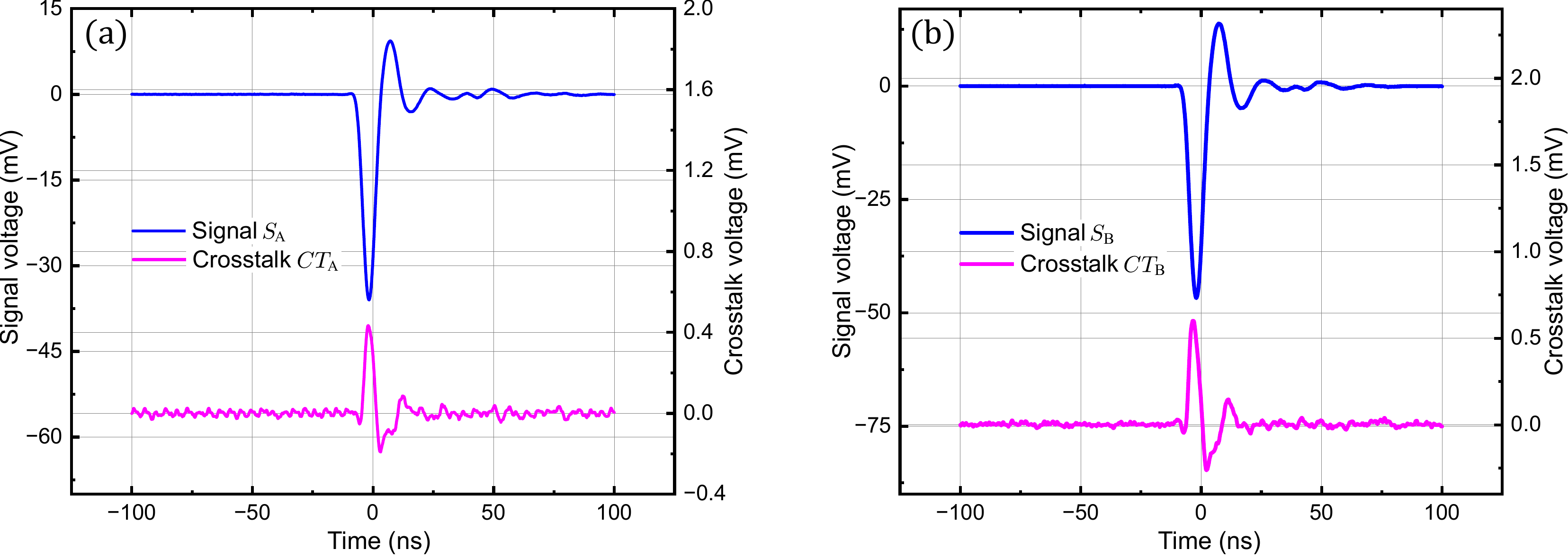}
        \caption{Background-subtracted time-averaged electron signal pulses from the CEMs and the resultant crosstalk on the start and stop circuits, respectively \textbf{(a)}, and vice versa \textbf{(b)}. The ratio of the height of the crosstalk pulse to that of the signal pulse was recorded to be $1.2\%$ (left) and $1.3 \%$ (right).}
        \label{fig:three graphs}
\end{figure*}

Fig.\ref{fig:three graphs} shows the CEM signal pulse on the start detector and the resulting crosstalk on the signal wire of the stop detector, as well as the signal pulse on the stop detector and the resulting crosstalk on the signal wire of the start detector, following background noise subtraction. The peak of the crosstalk pulses was measured to be $1.2\%$ and $1.3\%$ of the peak of the signal pulses, respectively.
\section{Elimination of crosstalk}

 We need to isolate a crosstalk-generated dip from a Coulomb or degeneracy pressure- generated one. To observe the reduction in coincidences due to crosstalk alone, we used a heated tungsten wire as a source of electrons. Thermally generated electrons are emitted continuously in time and from a larger surface area, so that Coulomb and degeneracy pressure are eliminated. A typical detection rate of $R=10^{5}$/s is low enough to ensure a long lifetime of the detectors. At that rate and for a detection efficiency of $\epsilon=10^{-2}$, emitted electrons are on average separated in time at the source by $100$ ns. Two electrons need to be within $\sim 10 $ fs to affect each other and cause a dip. Fig.\ref{fig:ZeroBounces}a shows the coincidence spectrum produced by this thermal random source. Still, a dip in coincidences is observed around the zero time delay. This is due to crosstalk. In the following section, we give the quantitative description of this dip. A Gaussian fit of this dip against the flat background indicates a maximum coincidence reduction of $8.5 \pm 0.4\%$ at $\tau=0$. In order to test the robustness of this crosstalk effect, the random source was replaced with a tungsten nanotip illuminated by a 800 nm pulsed laser with a pulse width of $\sim 400$ fs. For this long pulse duration, the electrons photoemitted by a single laser pulse are temporally separated to minimize their interaction. Fig.\ref{fig:ZeroBounces}b shows the coincidence spectrum generated by such a source. It consists of a series of peaks centered around integer multiples of $\tau_0$ as is expected for a pulsed electron source. The peaks were fitted with Gaussian functions to obtain the dashed red line curves in Fig.\ref{fig:ZeroBounces}b. The fit of the tungsten wire spectrum (red line in Fig.\ref{fig:ZeroBounces}a), is scaled to the average of the maximum heights of the Gaussian fits of the non-central peaks (red line in Fig.\ref{fig:ZeroBounces}b). The dip in this scaled spectrum, at a delay of $0\tau_0$, matches the fit to the experimental central-peak (in Fig.\ref{fig:ZeroBounces}b inset). This indicates that the antibunching is completely caused by the crosstalk. This approach offers a simple technique to correct for the effect of crosstalk in coincidence measurements. By using the normalized fit of a random continuous source and factoring it out from the measurement data, crosstalk can be eliminated. 

\section{Simulation of crosstalk}
\begin{figure}
     \includegraphics[width=8.5cm]{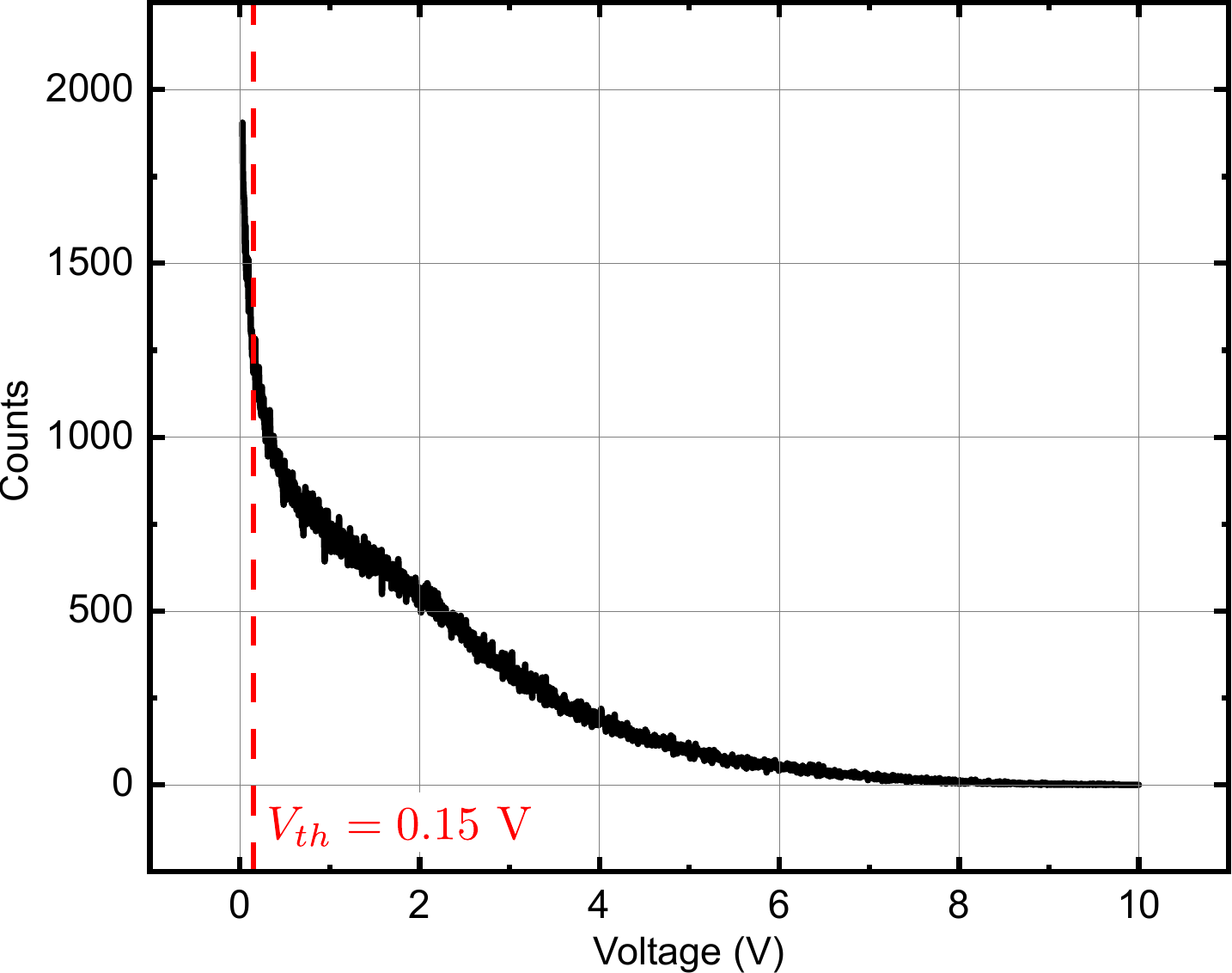}
     \caption{Pulse height distribution of the amplified electron signal pulses from the CEM. The red line indicates the discriminator setting $V_{th}=0.15$ V below which the signals are filtered out.}
     \label{fig:PHDstart}
\end{figure}

The coincidence spectrum is a histogram of the time delays observed between electron signals at the start and stop detectors. In our spectrum, the peaks have a FWHM of $4.73$ ns. This temporal spread is understood to arise from the uncertainty in time of arrival of the electrons from the source to the detector, different travel times of secondary avalanche electrons inside the CEM detectors, and the discriminator producing different delays while converting signal pulses to logical pulses. For the start detector signals, we model the temporal distribution as a Gaussian function $G_A(t)$ centered around $t=0$. For the stop detector pulses, we consider a series of Gaussian functions $G_B(t)=\sum_iG_B^i(t)$ where $G_B^i(t)$ are centered around the times $i\tau_0$; $i\in\{-3,...,3\}$, representing the temporal distribution of signals from an electron pulse train. The coincidence spectrum, $C(\tau)$, for non-interacting electrons unaffected by crosstalk can be mathematically represented as the cross-correlation between $G_A(t)$ and $G_B(t)$,
\begin{equation}
    C(\tau) = \int_{-\infty}^{\infty} G_A(t) 
     G_B(t + \tau) dt,
     \label{C(tau)}
\end{equation}
where a coincidence count at delay $\tau$ is generated when a start signal arrives at time $t$ and a stop signal arrives at $t+\tau$. When crosstalk is present, we can rewrite this equation for the coincidence spectrum as follows,
\begin{equation}
        C^{CT}(\tau) =  M_A^c(\tau)M_B^c(\tau)C(\tau).
    \label{crosstalk simulation}
\end{equation}
Here, $M_A^c(\tau)$ and $M_B^c(\tau)$ represent the average fractional loss of counts from the start and stop detectors, respectively. We define the pulse height as the extremum of the sum of the signal and the crosstalk, $M_A^s(\tau)$ and $M_B^s(\tau)$ as,
\begin{equation}
\begin{aligned}
 M_A^s(\tau)&=\text{min}_{t}(S_A(t)+CT_B(t+\tau)), \\
    M_B^s(\tau)&=\text{min}_{t}(S_B(t+\tau)+CT_A(t)).
\end{aligned}
\end{equation}
Here, the argument on the right-hand side is minimized with respect to t. The crosstalk in B, that is, $CT_A(t)$, occurs at the same time as the signal in A, $S_A(t)$, and vice versa. The measured signals and crosstalks are shown in Fig.\ref{fig:three graphs}. At $\tau=0$, $M_A^s(\tau)$ and $M_B^s(\tau)$ are the lowest since the minimum of the (negative) signal and the maximum of the (positive) crosstalk are aligned and they add up to give the most modified pulse height. Let us consider the average fraction of counts lost from the start detector at this point to be $F_{A}$ and that of the stop detector counts to be $F_{B}$. When $\tau$ is large enough so that the signal and crosstalk do not overlap, the fraction of counts lost is zero. We can use these two points to linearly extrapolate the lost fraction of counts $M_A^c(\tau)$ from $M_A^s(\tau)$,
\begin{equation}
    M_A^c(\tau)=\left(\frac{M_A^s(\tau)}{M_A^s(\infty)}-1\right)F_A+1,
    \label{M_A^c}
\end{equation}
and similarly for $M_B^c(\tau)$. This is a reasonable assumption for the small amount of crosstalk we are measuring in this experiment. We now proceed to find $F_{A}$ and $F_{B}$. The average number of counts measured at the start detector by sampling over all possible pulse heights of start and stop signals, $V_A$ and $V_B$, is proportional to,
\begin{equation}
        P_{A} \propto \int_{V_{th}}^\infty\int_{V_{th}}^\infty C_A(V_A)dV_AdV_B,
        \label{P}
\end{equation}
where, $C_A(V_A)$ denotes the pulse height distribution of the start signals (see Fig.\ref{fig:PHDstart}). The lower limit of the integral is set at the discriminator threshold voltage, $V_{th}$, below which the discriminators filter out the signals from both start and stop detectors. In the presence of crosstalk, the start and stop signal pulse heights decrease, which leads to a shift in the pulse height distributions. When the start and stop signal pulses align temporally, the average start detector counts measured with crosstalk, $P_{A}^{CT}$, is proportional to,
\begin{equation}
    P_{A}^{CT}\propto \int_{V_{th}}^\infty\int_{V_{th}}^\infty C_A(V_A+V_B^{CT})dV_AdV_B.
     \label{P_CT}
\end{equation}
Here, $V_B^{CT}$ is the height of the crosstalk pulse. This gives us the value of $F_A$ as,
\begin{equation}
    F_{A}=\frac{P_A-P_A^{CT}}{P_A}.
\end{equation}
Similarly, the proportionality constant for the stop signals $F_{B}$ can be measured taking the integrands as $C_B(V_B)$ in Eq.\eqref{P} and $C_B(V_B+V_A^{CT})$ in Eq.\eqref{P_CT}, where $C_B(V_B)$ denotes the pulse height distribution of the stop signals and $V_A^{CT}$ is the height of the crosstalk pulses. When we take $V_A^{CT}$ and $V_B^{CT}$ as $1.2\%$ of $V_A$ and $1.3\%$ of $V_B$ respectively, as we measured in our experiment (see Fig.\ref{fig:three graphs}), we get the value of $F_{A}$ and $F_{B}$ to be $3.3\%$ and $2.4\%$, respectively. We use these values of $F_{A}$ and $F_{B}$ in Eq.\eqref{M_A^c} to simulate the coincidence spectrum using Eq.\eqref{crosstalk simulation}. For a standard deviation of $1.41$ ns of the distributions $G_A(t)$ and $G_B^i(t)$, the width of the simulated coincidence peaks match those of the experimental ones. For a continuous source, the temporal distributions of the start and stop trigger pulses in Eq.\eqref{C(tau)} are replaced by unity. Fig.\ref{fig:ZeroBounces}a shows the simulated coincidence spectrum for a continuous thermal electron source (blue line). The shape of the simulated spectrum matches the measured spectrum in some detail. The magnitude of the central dip, $6.1\%$, differs from the experimental value, $8.5\%$. We attribute this discrepancy to the modified electronics while measuring the crosstalk with an oscilloscope in Fig.\ref{fig:three graphs}.

\section{Summary and Conclusion}
 Eq.\eqref{crosstalk simulation} uses as input the experimentally measured crosstalk, the signal voltage trace and the correction factor $F$. The factor $F$ in turn depends on the experimentally measured pulse height distributions and the set discriminator threshold, $V_{th}$, to predict the antibunching dip. This model explains about the $70\%$ of the experimental dip, while also providing understanding of the scaling of this crosstalk phenomenon. In addition, we find that shielding of the signal cables helps to minimize this pernicious effect.

To reduce the crosstalk further, increasing the threshold voltage, $V_{th}$, leads to a trade-off. At a higher threshold voltage, larger input pulses are accepted. This implies that the crosstalk pulses are larger as they are proportional to the input pulses. This leads to a larger percentage change in the signal counts, $P^{CT}$ (Eq.\eqref{P_CT}), and consequently in the number of coincidences as well (Eq.\eqref{crosstalk simulation}). On the other hand, a higher threshold could lead to a location in the pulse height distribution (Fig.\ref{fig:PHDstart}) that has a smaller slope, which leads to a smaller percentage change of $P^{CT}$. Note that the crosstalk dip remains, even for a flat pulse height distribution. Only a distribution with a local zero count value and a threshold set at that value, removes the crosstalk dip. Most CEMs and channel plates do not exhibit such a behavior.  
 
 The antibunching of electrons has been used to demonstrate the non-Poissonian statistics in previous experiments \cite{keramati2021,ropers2023,hommelhoff2023,luiten2024,hasselbach2002,kuwahara2021,kodama2011,hommelhoff2024}. Other correlation detectors such as the Timepix3 and other detector geometries may not suffer from the crosstalk problem discussed here. The search for free electron beams with quantum degeneracy that is distinguishable for Coulomb repulsion and relies on coincidence techniques and measurement of the correlation function $g^{(2)}$ is ongoing. Even new ideas to separate Coulomb from Pauli pressure are  being proposed \cite{hommelhoff2024}. In view of this, drawing attention to crosstalk may be of use. Keramati et. al. \cite{keramati2021} used a 800 nm 50 fs laser for photo-excitation and observed a reduction in coincidences by $\sim 24\%$ for a similar experimental setup as used here. The reduction was mostly attributed to Coulomb repulsion between electrons. We believe that it is likely that a part of the signal was due to crosstalk. Given that coincidence detection is used in a much broader category of experiments than addressed here, we hope that the model clarifies this type of crosstalk, which does not lead to spurious coincidences, but does lead to a loss of coincidence signal. Furthermore, and perhaps most importantly, we believe that measurements of the correlation $g^{(2)}$ for pulsed sources benefit from being accompanied by the same signal for a continuous source. The latter can be used for removing the crosstalk contribution to the anti-bunching spectrum. This may be especially relevant in view of the small expected quantum degeneracy signal.\\

\section*{Acknowledgments}
We gratefully acknowledge support by the U.S. National Science Foundation under Grant No. PHY-2207697. We thank Revanth Raghupatruni for the helpful discussions.
\bibliography{references.bib}

\end{document}